\documentclass[journal,transmag]{IEEEtran}

\usepackage{cite}
\usepackage{graphicx}
\usepackage{amssymb,amsmath}

\begin{document}

\title{Machine Learning Quantum Systems with Magnetic p-bits
}


\author{
	\IEEEauthorblockN{Shuvro Chowdhury\IEEEauthorrefmark{1,$\circ$},~\IEEEmembership{Member,~IEEE},  Kerem Y. Camsari \IEEEauthorrefmark{1,$\dagger$}~\IEEEmembership{Senior Member,~IEEE}}

	\IEEEauthorblockA{\IEEEauthorrefmark{1}Department of Electrical and Computer Engineering, University of California, Santa Barbara, \\ Santa Barbara, CA, 93106, USA\\ $^\circ$schowdhury@ece.ucsb.edu, $^\dagger$camsari@ece.ucsb.edu}
}

\IEEEtitleabstractindextext{%
\begin{abstract}

\noindent The slowing down of Moore's Law has led to a crisis as the computing workloads of Artificial Intelligence (AI) algorithms continue skyrocketing. There is an urgent need for scalable and energy-efficient hardware catering to the unique requirements of AI algorithms and applications. In this environment, probabilistic computing with p-bits emerged as a scalable, domain-specific, and energy-efficient computing paradigm, particularly useful for probabilistic applications and algorithms. \vspace{3pt}

\noindent In particular, spintronic devices such as stochastic magnetic tunnel junctions (sMTJ) show great promise in designing integrated p-computers. Here, we examine how a scalable probabilistic computer with such magnetic p-bits can be useful for an emerging field combining machine learning and quantum physics. 


\end{abstract}

}

\maketitle

\pagestyle{empty}
\thispagestyle{empty}

\IEEEpeerreviewmaketitle

\section{Introduction}


\IEEEPARstart{W}{ith} the slowing down of Moore’s Law, initial efforts to ``re-invent the transistor''  met
significant challenges \cite{theis2010s}. An emerging idea in the field has been to augment the capabilities of the CMOS transistor, instead of finding a drop-in replacement in integrated CMOS + X architectures. Here, we describe how CMOS + spintronics technology, in particular, stochastic magnetic tunnel junctions can be an energy-efficient building block to solve the quantum many-body problems with probabilistic computers.

The idea of a probabilistic computer with probabilistic building blocks can be tied back to a famous talk given by Feynman \cite{feynman1982simulating} widely credited for starting the field of quantum computation. In this talk, Feynman also described a probabilistic computer with the same basic idea that mapping a problem to the evolution of a physical system allows energy-efficient, natural computation. In the last few years, the concept of building such a probabilistic computer with p-bits \cite{camsari2017stochasticL} has attracted significant attention with a wide range of applications from combinatorial optimization \cite{aadit2022massively} to machine learning \cite{kaiser2022hardware}. 
\section{Spintronics for Probabilistic Computation}

\begin{figure}[t!]
\vspace{0pt}
\centering
\includegraphics[keepaspectratio,width=0.99\columnwidth]{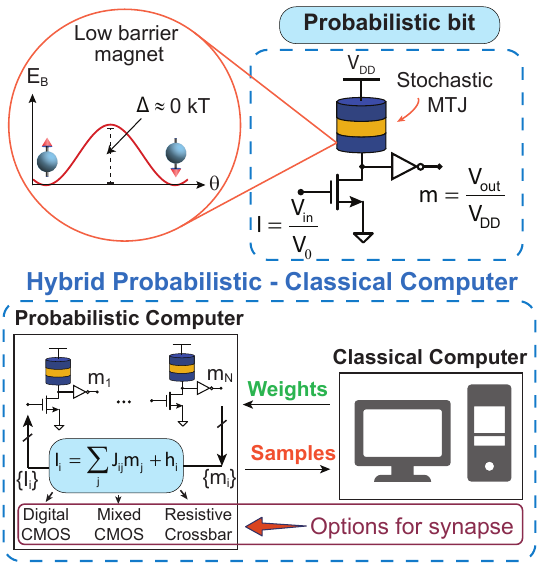}\\
\caption{
(Top) Stochastic magnetic tunnel junction-based p-bit. The bottom layer of the MTJ has been modified to a low-barrier ($\sim 0$ kT) magnet to achieve the stochastic behavior. (Bottom) A hybrid probabilistic-classical computer where the p-computer solves the computationally hard problem of obtaining probabilistic samples from a given set of weights and the classical computer finds the new weights based on the provided samples. }
\label{fig:figmagnet}
\vspace{-15pt}
\end{figure}

There are many possible physical implementations of probabilistic bits: digital approximations using pseudo-random number generators \cite{aadit2022massively}, single photon avalanche diodes \cite{whitehead2022cmos}, diffusive memristors \cite{woo2022probabilistic} and other naturally noisy devices.  We believe, however, that spintronic technology in the form of CMOS-integrated magnetic tunnel junctions holds the greatest promise in designing scalable p-computers to transform machine learning and AI applications. 

Designing magnetic tunnel junctions out of low-energy barrier nanomagnets (LBM) allows a steady stream of noisy bits in highly compact circuit topologies. In recent years, the potential of fast fluctuations (down to nanoseconds) using sMTJs with in-plane magnetic anisotropy has been experimentally demonstrated \cite{hayakawa2021nanosecond,safranski2021demonstration} following theoretical expectations \cite{kaiser2019subnanosecond}. The sheer scalability of CMOS-integrated MTJ technology, along with extremely compact p-bits operating at room temperature put them in a unique position for the development of highly scalable and energy-efficient p-computers. Fig.~\ref{fig:figmagnet} shows the general architecture of such a p-computer combining stochastic magnetic tunnel junctions (p-bits), coupled by a ``synapse'' circuit to solve a wide variety of problems relevant to machine learning and AI. Just as there are different options for physical p-bits, several different options to implement synapses exist (Fig.~\ref{fig:figmagnet}). \clearpage

\vspace{-5pt} 
\section{Machine learning quantum systems with p-bits}
\label{sec:sec3}
\vspace{-3pt} 

\begin{figure}[t!]
\vspace{0pt}
\centering
\includegraphics[keepaspectratio,width=0.9\columnwidth]{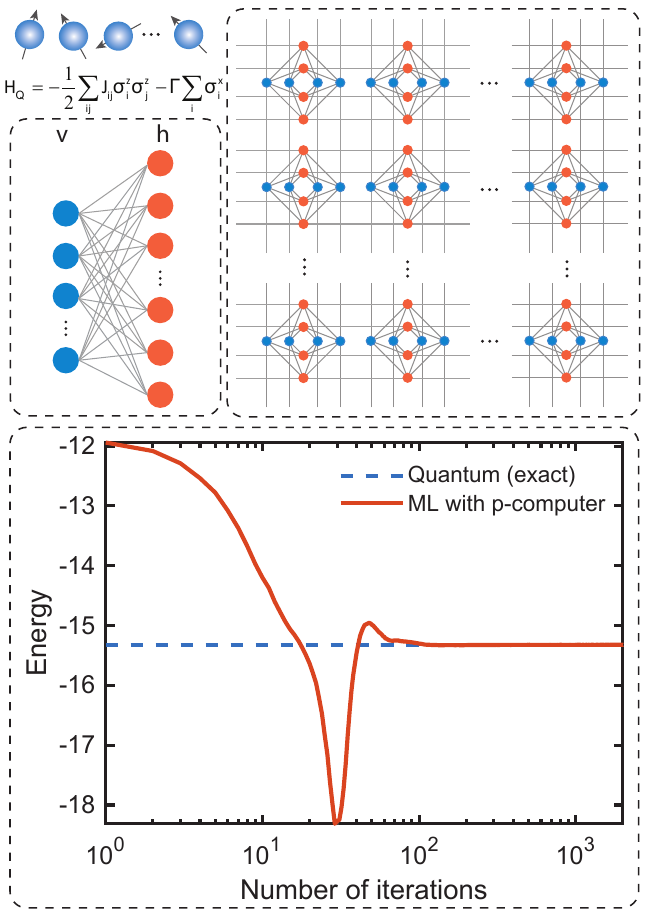}\\
\caption{\textbf{Machine learning quantum systems with p-bits}:  (Top-left) A transverse field ($\Gamma = +1$) is applied to a linear chain of twelve ($12$) FM coupled ($J_{ij} = +1$) qubits with periodic boundary conditions.  (Middle-left) To obtain the ground state of this quantum system a restricted Boltzmann Machine (RBM) is employed with 12 visible and 48 hidden nodes where all nodes in the visible layer are connected to all nodes in the hidden layer. (Top-right) this machine learning model is then embedded onto a hardware amenable sparse p-bit network arranged in a chimera graph using minor graph embedding. We use a coupling strength of 1.0 among the replicated visible and hidden nodes in the embedded p-bit network. (Bottom) The FPGA emulation of this probabilistic computer performs reinforcement learning in tandem with a classical computer and gradually learns the quantum (exact) result as shown.}\label{fig:ML}\vspace{-15pt} \end{figure}

One of the most anticipated applications of quantum computers is simulating the behavior of many-body quantum systems. Following the success of machine learning and AI algorithms, a promising recent approach has been to train stochastic neural networks (such as Boltzmann Machines) to approximately solve the quantum many-body problem by a variational guess \cite{carleo2017solving}. This approach is primarily implemented in high-level software implementations which dictate the model sizes and choices. For example, the use of restricted Boltzmann Machines (RBM) over the more powerful unrestricted or deep BMs is motivated by the former's efficient software implementation in synchronous systems. With scaled p-computers using millions of magnetic p-bits, massively parallel and energy-efficient \textit{hardware} implementations of the much more powerful unrestricted/deep BMs become feasible, paving the way to simulate practical quantum systems. 

As a representative example of this approach, we show how p-bits laid out in sparse, hardware-aware graphs can be used for machine learning quantum systems (Fig.~\ref{fig:ML}). The target problem is to find the ground state of a many-body quantum system, in this case, a 1D-Transverse Field Ising Hamiltonian, a typical spin system that can be considered the quantum generalization of the classical Ising Hamiltonian. The ground state wavefunction of this system is expressed as a variational guess over parameters of a probabilistic neural network,  which, in this simple example is an RBM. Then, with a combination of probabilistic sampling and weight updating, the variational guess is adjusted such that the parameters of the RBM point to the ground state of the desired quantum Hamiltonian. 

The approach to implementing this scheme with p-bits differs at this stage. First, the RBM with its dense (all-to-all) connections between the visible and hidden layers is not conducive to being implemented in massively parallel, scalable p-computers. For this reason, one approach is to map the RBM to a sparse graph, such as the Chimera lattice, using a process called minor graph embedding. Then, using a hybrid setup with fast sampling in a probabilistic computer coupled to a classical computer, the iterative process of sampling and weight updating can be performed. Having a massively parallel and fast sampler using scaled p-computers with magnetic p-bits allows the selection of higher-quality states of the wavefunction to update the variational guess. Fig.~\ref{fig:ML} shows an example simulation of how a p-computer learns the ground state of a 1D TFIM model. The scaling of p-computers using magnetic p-bits may allow much larger implementations of quantum systems in the future. \vspace{3pt}


This work was supported through NSF (CAREER) CCF-2237357 and ONR YIP grants.

\appendix
\noindent \textbf{Details of the machine learning quantum systems with p-bits example: } In the example shown in Section \ref{sec:sec3}, we find the ground state energy of $N$ spin nearest neighbor 1D transverse field Ising model although the method used in this paper is fully general to be applied to non-nearest neighbor models. The Hamiltonian under discussion is then written as follows:
\begin{equation}
H_Q = -\left(\sum_{i}^{N} J_{i,i+1}\sigma_i^z\sigma_{i+1}^z+\Gamma\sum_{i}^{N}\sigma_i^x\right)
\end{equation}
where $\sigma_i^{z}=I^{\otimes i-1}\sigma^z I^{\otimes N-i}$ with $I$ and $\sigma^z$ being the $2\times2$ identity matrix and Pauli-Z matrix respectively. $\sigma_i^{x}$ is defined in a similar manner. $J_{i,i+1}$ is the coupling strength between qubit $i$ and $i+1$ and $\Gamma$ is the strength of the transverse field. We also use a periodic boundary at the ends i.e., $J_{N,N+1} = J_{N,1}$. 

In this work, we used 12 spins, i.e., $N=12$, we also set $\Gamma = 1.0$, and $J_{i,i+1} = 1.0$ for all $i$. In this case, the ground state energy of the problem is known to be $E_0= 15.32256$.

In order to solve this problem with ML techniques, we use an RBM with 12 visible nodes and $\alpha = 4$ ($\alpha$ is the ratio of the number of hidden and visible nodes) following the suggestion in \cite{carleo2017solving}. However, with this model, each visible node has 48 connections to connect with all hidden nodes, and as mentioned in the text, it is very difficult to implement such a large number of connections in hardware. Therefore with the hardware implementation in mind, we map this RBM model into a Chimera graph where the basic unit is a $4\times4$ RBM (4 visibles and 4 hidden nodes) and use the minor graph embedding (MGE) technique from DWAVE [citation required]. At the end of MGE, the above-mentioned RBM model with 12 visible and 48 hidden nodes is mapped into a $(12,3,4)$ Chimera graph with 12 rows and 3 columns of $4\times4$ RBM units. This requires in total $12\times3\times8=288$ p-bits in total and we set the coupling strength to $1.0$ for connections which connect the copies of a given visible/hidden node. The other advantage of using the Chimera graph is that each p-bit now has only 6 neighbors and therefore only 6 wires to process. 

We then synthesize this mapped network in Xilinx Alveo U250, a data center accelerator card (Virtex UltraScale+ XCU250 FPGA) with peripheral component interconnect express (PCIe) connectivity \cite{xilinx-u250}. PCIe interface performs data transfer at the rate of 2.5 gigatransfers per second (GT/s). The classical computer used in this study is equipped with an 11\textsuperscript{th} Gen Intel Core i7-11700 processor with a clock speed of up to 4.90 GHz and 32 GB of random access memory (RAM). To strike a balance between accuracy and scalability, in FPGA, we set the precision to s{6}{3} (fixed-point precision: 1 bit for sign, 6 bits for integer, and 3 bits for fraction) instead of using floating point precision as one would use in general purpose processors but requires more resources per p-bit in FPGA.

\begin{figure}[t!]
\vspace{0pt}
\centering
\includegraphics[keepaspectratio,width=0.85\columnwidth]{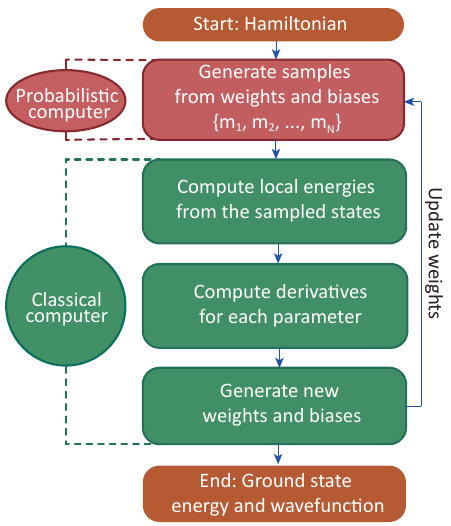}\\
\caption{\textbf{Machine learning quantum system with hybrid scheme flowchart}: The probabilistic computer generates and sends samples to the general purpose processor (CPU) based on the weights and biases received from the latter. Based on the samples received, the 
 CPU computes the local energies and the corresponding derivatives and then computes the new weights and biases which are then sent back to the probabilistic computer until a convergence is reached.}\label{fig:MLAlgo}\vspace{-20pt} \end{figure}

The machine learning of the quantum problem is then performed in a hybrid manner where a CPU provides the weights and biases for the synthesized Chimera network and the network then produces ultra-fast samples from the corresponding Boltzmann distribution. These samples are then transferred back to the CPU which then computes the local energies, and derivatives and updates the weights and biases. The process flow is shown in Fig. \ref{fig:MLAlgo}.

The minor graph embedding approach usually increases the number of p-bits required more than the original size of the problem. This can be avoided by changing the topology from RBM which has full connectivity between the two layers to a \emph{further} restricted BM (FRBM) which has limited connectivity per p-bit between two adjacent layers. The other possibility is to use a different sparse graph other than RBM. But this approach will also require finding the analytical derivative for that graph or deploying numerical methods to compute the derivatives on the fly.

Our current approach of using a probabilistic computer cuts down the cost of generating samples on the CPU to a negligible value and improves the time per epoch by approximately a factor of two. However, the dominating bottleneck in our present implementation is the computation of local energies whose complexity grows as $O(n^2)$ on CPUs, $n$ being the problem size. One possibility is to compute local energies on the fly inside the FPGA,  alleviating this bottleneck. For rapid inference where the goal is to successively evaluate a trained wavefunction, our present implementation should provide significant speedup. The possibility of performing training and/or inference of neural quantum states with direct hardware implementations of Boltzmann Machines (restricted and unrestricted \cite{niazi2023training}) is an intriguing new direction to classically simulate quantum systems. 

\end{document}